\documentstyle[psfig]{l-aa}

\begin{document}

   \thesaurus{06         % A&A Section 6: Form. struct. and evolut. of stars
              (03.11.1;  % Cosmogony,
               16.06.1;  % Planets and satellites: general,
               19.06.1;  % Solar system: general,
               19.37.1;  % Stars: formation of,
               19.53.1;  % Stars: oscillations of,
               19.63.1)} % Stars: structure of.
   \title{An optical and near IR study of the old open cluster
   NGC~2141\thanks{Based on observations taken at ESO La Silla and TIRGO.}}

   \author{G. Carraro\inst{1}, S. M. Hassan\inst{2},  S. Ortolani\inst{1} 
          and A. Vallenari\inst{3} 
		  }

   \offprints{G. Carraro ({\tt carraro@pd.astro.it})}

   \institute{Dipartimento di Astronomia, Universit\'a di Padova,
	vicolo dell'Osservatorio 5, I-35122, Padova, Italy
        \and
National Research Institute of Astronomy and Geophysics,
   	Helwan, Cairo, A.R.E.
   		\and
        Osservatorio Astronomico di Padova, vicolo Osservatorio 
          5, I-35122, Padova,
	Italy\\
        e-mail: {\tt 
carraro,ortolani,vallenari\char64pd.astro.it}
             }

   \date{Received ; accepted}

   \maketitle

   \markboth{Carraro et al}{NGC~2141}

   \begin{abstract}
We report on  CCD optical ($B$ and $V$ passbands)
and near IR ($J$ and $K$ bands) observations in the region of the old
open cluster NGC~2141. By combining the two sets of photometry 
(500 stars in common) we
derive new estimates of the cluster fundamental parameters.
We confirm that the cluster is 2.5~Gyrs old, but, with respect to
previous investigations, we obtain a slightly larger reddening
(E(B-V)~=~0.40), and
a slightly shorter distance (3.8~kpc) from the Sun. Finally we
present Luminosity Function (LF)
in the V band, which is another
age indicator. We provide a good fit for the age range inferred
from isochrones by assuming the Kroupa et al. (1993) IMF up
to $M_V$~=~5.0. We interpret the disagreemt at fainter magnitudes
as an evidence of mass segregation.

   \end{abstract}

%
%  14.Sep.'90: Demo-Vs.
%________________________________________________________________

\begin{figure*}
\centerline{\psfig{file=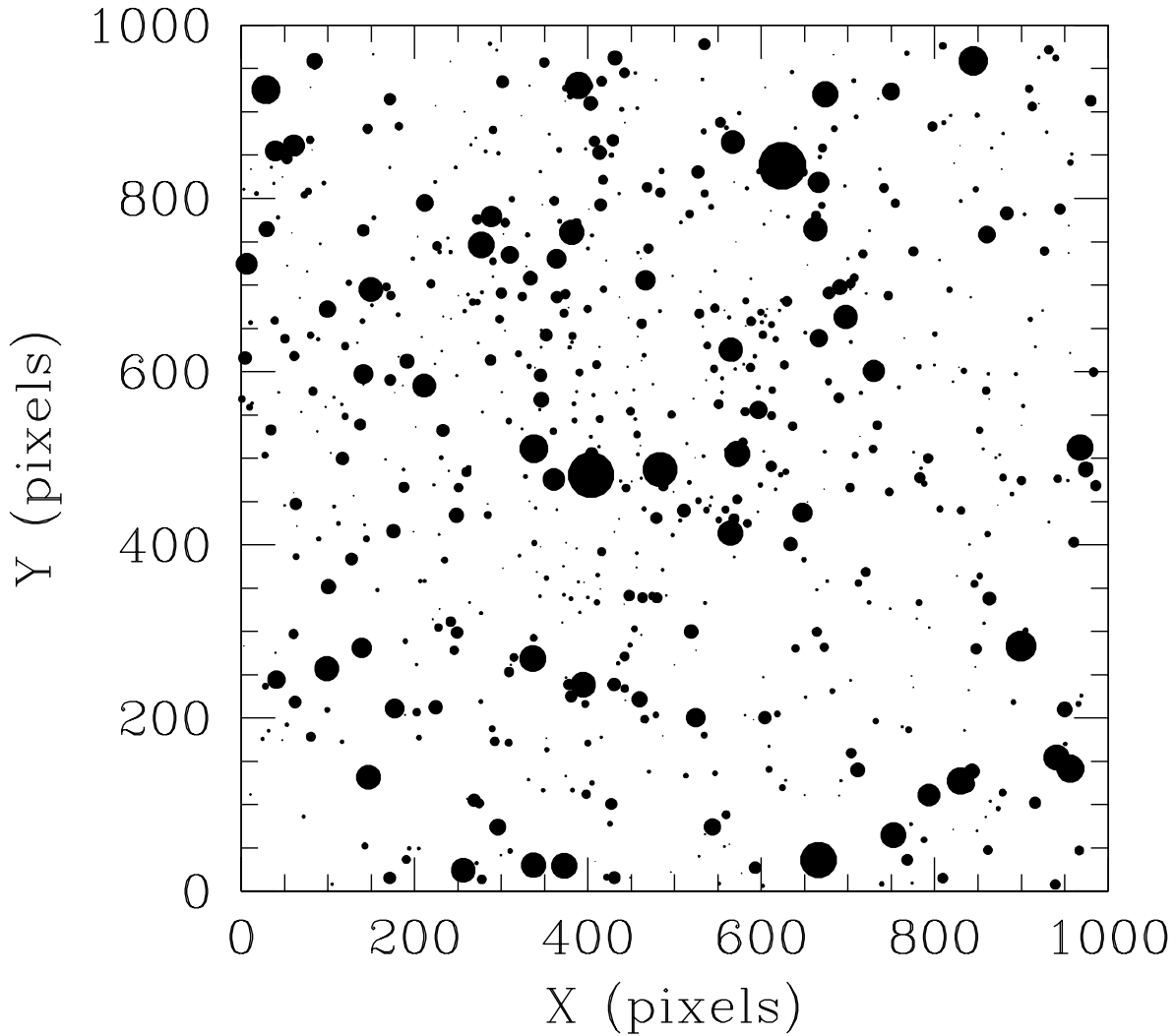,width=16cm,height=16cm}}
\caption{The field covered in the region of NGC~2141 by the optical
photometry. The mosaic
of the four fields observed in the IR covers almost the same
region. North is up, East on the left.}
\end{figure*}

\section{Introduction}
In this paper we continue a 
series dedicated to the presentation of near-infrared 
photometry (in $J$ and $K$ pass-bands) for northern galactic open clusters.
We already reported on the very young open clusters NGC~1893 and Berkeley~86 
(Vallenari et al 1999), the old clusters Berkeley~17 and Berkeley~18 
(Carraro et al 1999a), the intermediate age clusters
IC~166 and NGC~7789 (Vallenari et al 2000) and King~5 
(Carraro \& Vallenari 2000). 
Here we combine optical $B$ and $V$ and near-infrared $J$ and $K$ photometry for
NGC~2141, a faint 
old open cluster which did not  receive much  attention 
in the past.\\
NGC~2141 is located  close to the galactic plane in the antic-enter direction
at $l~=198~^{o}.75$ and $b~=-5^{o}.79~$, and it is designated also as OCL~487
(Lund~203) and C0600+104 by IAU. 
Its diameter is estimated to be 
about $10^{\prime}$.\\
A preliminary investigation was conducted by Burkhead et al (1972) who obtained
photographic and photoelectric $UBV$  
photometry for about 300 stars with the aim of assessing whether the cluster
is very old.
This study revealed that NGC~2141 is a cluster of late intermediate age,
4.4~kpc distant from the sun, and with a reddening E(B-V)=0.30.\\
The metal abundance of NGC~2141 has been 
determined several times in the past.
Janes (1979) obtained a value $[Fe/H]~=~-0.54\pm0.42$ from DDO photometry,
while Geisler (1987) obtained  $[Fe/H]~=~-0.63\pm0.15$ from Washington photometry.
Finally, by using medium resolution spectroscopy of 
six giant stars Friel \& Janes (1993) found $[Fe/H]~=~-0.39\pm0.11$

The kinematics of NGC~2141 has been studied by Friel et al (1989) and Friel (1989).
This latter study (5 stars) indicates  that the radial velocity is $V_r ~=~43\pm6~km/s$.
Individual radial velocities for 15 stars have been obtained 
by Minniti (1995). This survey has only one star in common with Friel (1989)
and the radial velocity estimate is in agreement, suggesting that most of
Minniti (1995) stars might  probably be cluster non members. Anyway, a much 
deeper investigation is required to isolate cluster members in NGC~2141.\\
More recently Rosvick (1995) published optical $VI$ photometry of 3561 stars
in NGC~2141 together with $JH$ infrared photometry for a handful of stars.
This is the first comprehensive study of NGC~2141. The author
infers by comparison with half solar
(Z~=~0.006) isochrones an age of 2.5~Gyr, a distance 
of 4.2~kpc and a reddening E(B-V)=0.35.\\

In this paper we combine IR $JK$ (765 stars)
and optical $BV$ (1073 stars) photometry to obtain
new estimates of the cluster fundamental parameters.
The layout of the paper is as follows: Section~2 is devoted to the presentation
of data acquisition and reduction; Section~3 deals with the morphology
of the Color Magnitude Diagrams (CMDs) for different pass-bands;
Section~4 concerns the derivation of the cluster metallicity,
whereas Section~5 deals with the estimate of color excess.
Section~6 is devoted to infer the age and distance, while
in Section~7 we discuss the Luminosity Function. Finally our conclusions
are summarized in Section~8.

\begin{table*}
\caption[ ]{ TIRGO observation Log Book }
\tabcolsep 0.8cm
\begin{tabular}{c|c|c|c|c|c|c}
\hline
\hline
Cluster               &$\alpha$    &$\delta$  & Date& \multicolumn{2}{c}
{Exposure Times (sec)} & Field of view\\ 
                           &(2000)      &(2000) &    &J&K & \\
\hline
NGC~2141  & 06 02 58.2& 10 26 39& Oct, 23, 1997& 480 & 680 & $7^{\prime}.5 \times 7^{\prime}.5$\\
\hline
\hline
\end{tabular}
\end{table*}

\begin{table*}
\caption[ ]{2.2m ESO/MPI observation Log Book }
\tabcolsep 0.8cm
\begin{tabular}{c|c|c|c|c|c|c}
\hline
\hline
Cluster               &$\alpha$    &$\delta$  & Date& \multicolumn{2}{c}
{Exposure Times (sec)} & Field of view\\ 
                           &(2000)      &(2000) &    &B&V &\\
\hline
NGC~2141  & 06 02 58.2& 10 26 39& Dec, 8, 1991& 1200 & 4200 & $5^{\prime}.7 
\times 5^{\prime}.7$\\\\
\hline
\hline
\end{tabular}
\end{table*}

\section {Observations and Data reduction}

\subsection{TIRGO observations}
J (1.2 $\mu$m) and K (2.2 $\mu$m) photometry of NGC~2141
was obtained 
at the  1.5m Gornergrat Infrared Telescope (TIRGO) 
equipped with Arcetri Near Infrared Camera (ARNICA)
in October 1997. 
ARNICA is based  on a NICMOS3 256$\times$ 256
pixels array (gain=20 e$^-/$ADU, read-out noise=50 e$^-$,
 angular scale =1$\arcsec/$pixel, and $4^{\prime}  \times 4^{\prime}$ field of
view). 
More details about the observational equipment and infrared camera,
and the reduction procedure can be found in Carraro et al (1999a).
Through each filter 4 partially overlapping images of
each field were obtained,
 covering a total field of view of about $7^{\prime}.5 \times 7^{\prime}.5$,
in short exposures to avoid sky saturation.

 The log-book of the observations is presented
 in Table~1 where the center of the observed field and the total
exposure times are given.
 The night was photometric
with a seeing of 1$\arcsec$-1.5$\arcsec$. 
Fig.1 presents the final mosaic of the 4 frames
for NGC~2141 in K passband.

The conversion of the
instrumental magnitude j and k to the standard
J, K was made using stellar fields of standard stars taken
from   Hunt et al (1998) list.
About 10 standard stars per night have been used. 

The  relations in usage per 1 sec exposure time are: 

\begin{equation}
J  = j+19.51 + k_J \times 1.03
\end{equation}
\begin{equation}
K  = k+18.94 +k_K \times 1.06
\end{equation}

where $k_J$ and $k_K$ (the extinction coefficients, in magnitude per airmass)
are 0.25 and 0.10, respectively.
The standard deviation of the zero points of 0.03  mag for the $J$ 
and 0.04 for the $K$ magnitude. This error is only due to the linear
interpolation of the standard stars. The calibration uncertainty
is dominated by the error due to the correction from aperture photometry
to PSF fitting magnitude. 
The standard stars used for the calibration do  not cover the entire
color range of the data, because of the lack of stars redder than
$(J-K) \sim 0.8$. From our data, no color term is found for $K$ mag,
whereas we cannot exclude 
it  for the $J$ magnitude.
Taking all into account,
we estimate that the total error
on the calibration is about 0.1 mag in both $J$ and $K$ pass-bands
(see Vallenari et al 2000 for additional details).

\subsection{2.2m ESO/MPI observations}
NGC~2141 was observed with the 2.2m ESO/MPI telescope at La Silla. The
focal reducer ESO EFOSC 2 camera was used, equipped with the Thompson UV coated 
1000 $\times$ 1000 pixels CCD ESO $\#$~19. The chip has 19 micron pixel size,
corresponding to $0^{\prime\prime}.34/pixel$ projected on the sky.
The total field of view is about $5^{\prime}.7 \times 5^{\prime}.7$ .
The observations have been carried out in 1991 December 8 (see Table~2
for details.) The night was photometric with an average seeing of 
$1^{\prime\prime}.5$.
Several standard field stars from Landolt (1992) have been observed
during the same night, from which the following color equations
have been derived:

\[
V = v + 25.75 + 0.05 \times (B-V)
\]

\[
B = b + 26.11 +0.21 \times (B-V)
\]

After the standard pre-processing, instrumental magnitudes have 
been extracted using DAOPHOT~II and the accompanying ALLSTAR program
in the MIDAS environment. Since the field is relatively crowding free,
the aperture corrections have been directly computed on the original
frames, giving the coefficient within a 0.02 magnitude spread.
The final errors in the zero points amount to 0.02 both in $B$ and in $V$.
The comparison of 5 stars in common with Burkhead et al (1972) 
yields:

\[
V_{BBH} -V_{CHOV} = 0.01\pm0.01
\]

\[
(B-V)_{BBH} -(B-V)_{CHOV} =0.02\pm0.02
\]

where $BBH$ and $CHOV$ refer to Burkhead et al (1972) and the present work,
respectively.\\

We estimated the photometric errors by means of experiments with artificial stars
(Carraro \& Ortolani 1994), obtaining errors of 0.02, 0.04 and 0.08 at $V$ = 18,
19 and 20 mag, respectively. Another estimate of the photometric errors
derives from the MS natural width. At the same above magnitude levels we found 
dispersions in color of 0.11, 0.16 and 0.20. These latter values clearly take 
into account also the effect of unresolved binaries 
(see the sequence rightwards of the MS in Figs.~2 and 3) and possible 
variable reddening.

\begin{figure}
\centerline{\psfig{file=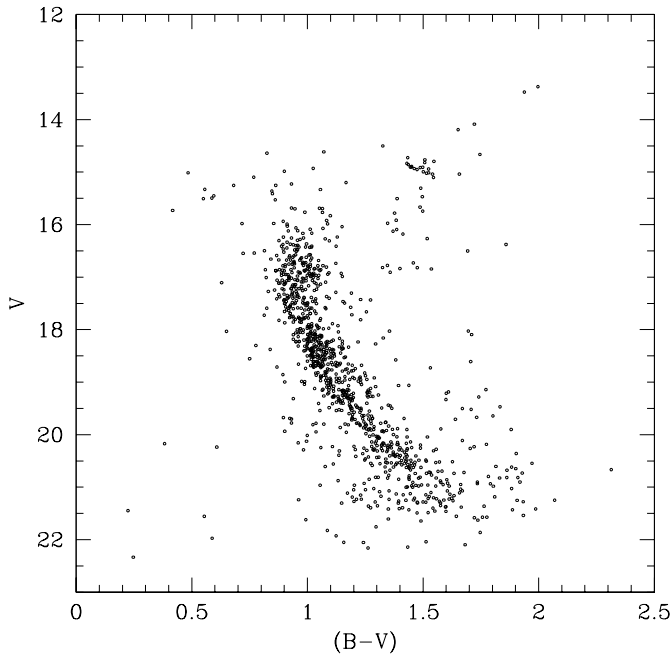,width=9cm,height=9cm}}
\caption{The CMD of NGC~2141 derived from optical photometry.}
\end{figure}

\begin{figure}
\centerline{\psfig{file=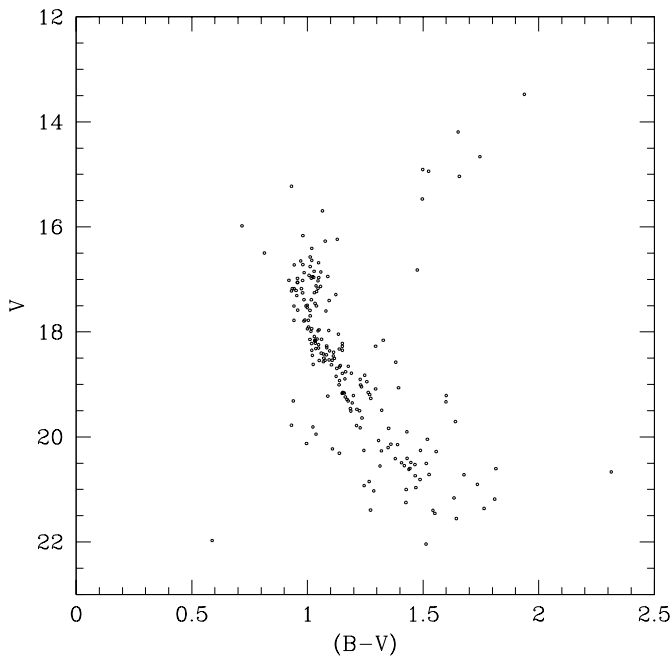,width=9cm,height=9cm}}
\caption{The CMD of NGC~2141 derived from optical photometry
by considering only stars within 1.2 arcmin from the cluster center
. Note the presence of a parallel sequence red-wards to the MS, which 
we ascribe to the presence of unresolved binaries}
\end{figure}

\begin{figure}
\centerline{\psfig{file=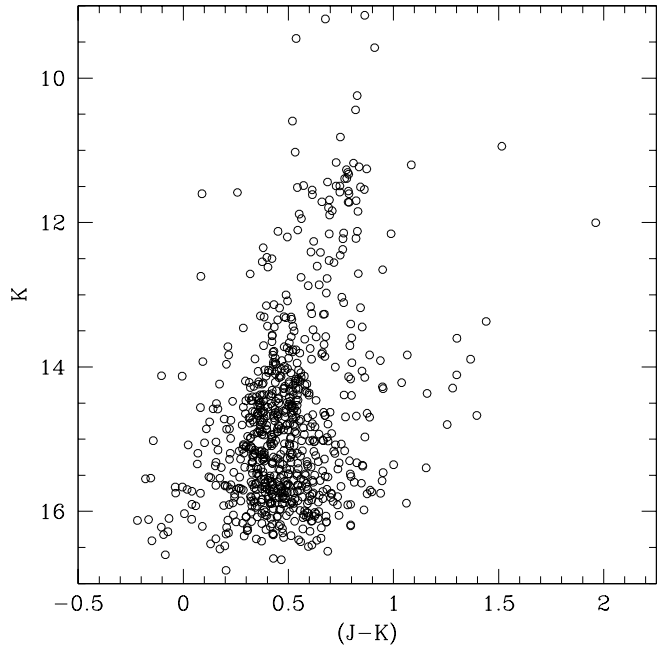,width=9cm,height=9cm}}
\caption{The CMD of NGC~2141 derived from near IR photometry}
\end{figure}

\section{The Color Magnitude Diagram}
In this section we discuss separately the CMDs obtained
from the optical and near IR photometry.

\subsection{The optical CMD}
The CMD derived from optical photometry is shown in Fig.~2.\\
The global morphology resembles the CMD of an intermediate age or old open cluster
like NGC~7789 (Girardi et al 2000a), IC~4651 (Bertelli et al.  1992)
and NGC~2204 (Kassis  et al. 1997), say a cluster whose age is between
NGC~752 (1.7 Gyr) and M~67 (4.0 Gyr) (Carraro et al. 1999b). 
The MS extends down to $V \approx
22$, and the Turn Off point (TO) is situated at $V~=~15-16.5$, $(B-V)~=~0.9$,
with a few star spreading towards brighter regions ($V \approx 16$).
The concentration  of stars in the red region of the diagram  
(at $V~\approx~15.0$, $(B-V)~\approx~1.5$)
represents the Red Giant (RG) clump
of core He-burning stars. Note the tilt and the extension
of the clump in color,
which can be ascribed to a possible spread in metallicity or in age, or
to the presence of some differential reddening.
The most reasonable explanation is a metallicity effect. Indeed the MS
is rather thin up to the limiting magnitude, thus ruling out a
significant age dispersion.
A secondary sequence of unresolved binary stars is visible on the red
side of the MS. This is more evident from Fig.~3,
where only the core of the cluster is considered.
By counting the number of stars belonging to the two different sequences,
we find that the binary percentage amounts as minimum at 30\%. 
Finally,  the Herzsprung gap is clearly defined, while the sub-giant
and RG branch are scarcely populated.\\
The population of stars on the right side of the MS and above the TO 
probably belongs to the field or they might be blue
stragglers which are members of the cluster.
The field stars contamination in fact is not very
high. The MS is much thinner than in Rosvick (1995)
even after the field stars decontamination (see her Fig.~5). 
This rules out the
presence of a significant differential reddening across the $5^{\prime}
\times 5^{\prime}$ region covered by the present photometry.\\

\subsection{The near IR CMD}
The CMD from IR photometry is shown in Fig.~4.
The MS extends down to $K~=~16.5$, and the TO is
located at $K\approx14.5$, $(J-K)\approx0.40$. The region above the TO 
might contain some interlopers (see Fig.~2).\\
The stars at $K \approx 11.5$, $(J-K) \approx 0.8$
might represent the clump of core He burners.
The cluster is indeed faint, and the MS gets rather
wide down to $K~\approx 15$. The MS is much larger than
in the optical CMD, due to the larger photometric errors 
we have in the IR photometry (see Section~2).
Apparently there is no Herzsprung gap,
and the RGB is scarsely populated, but sufficiently evident
to be used to estimate cluster metallicity (see next Section).

\begin{figure}
\centerline{\psfig{file=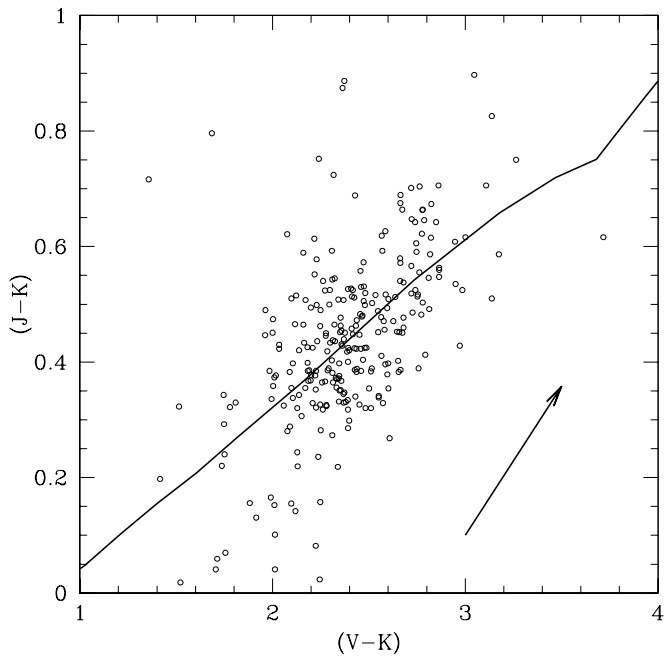,width=9cm,height=9cm}}
\caption{Two colors diagram for MS stars in NGC~2141. The solid line is a Zero
Age MS for $[Fe/H]~=~-0.43$. In the lower right corner the reddening vector is shown.
See text for any detail.}
\end{figure}

\begin{figure}
\centerline{\psfig{file=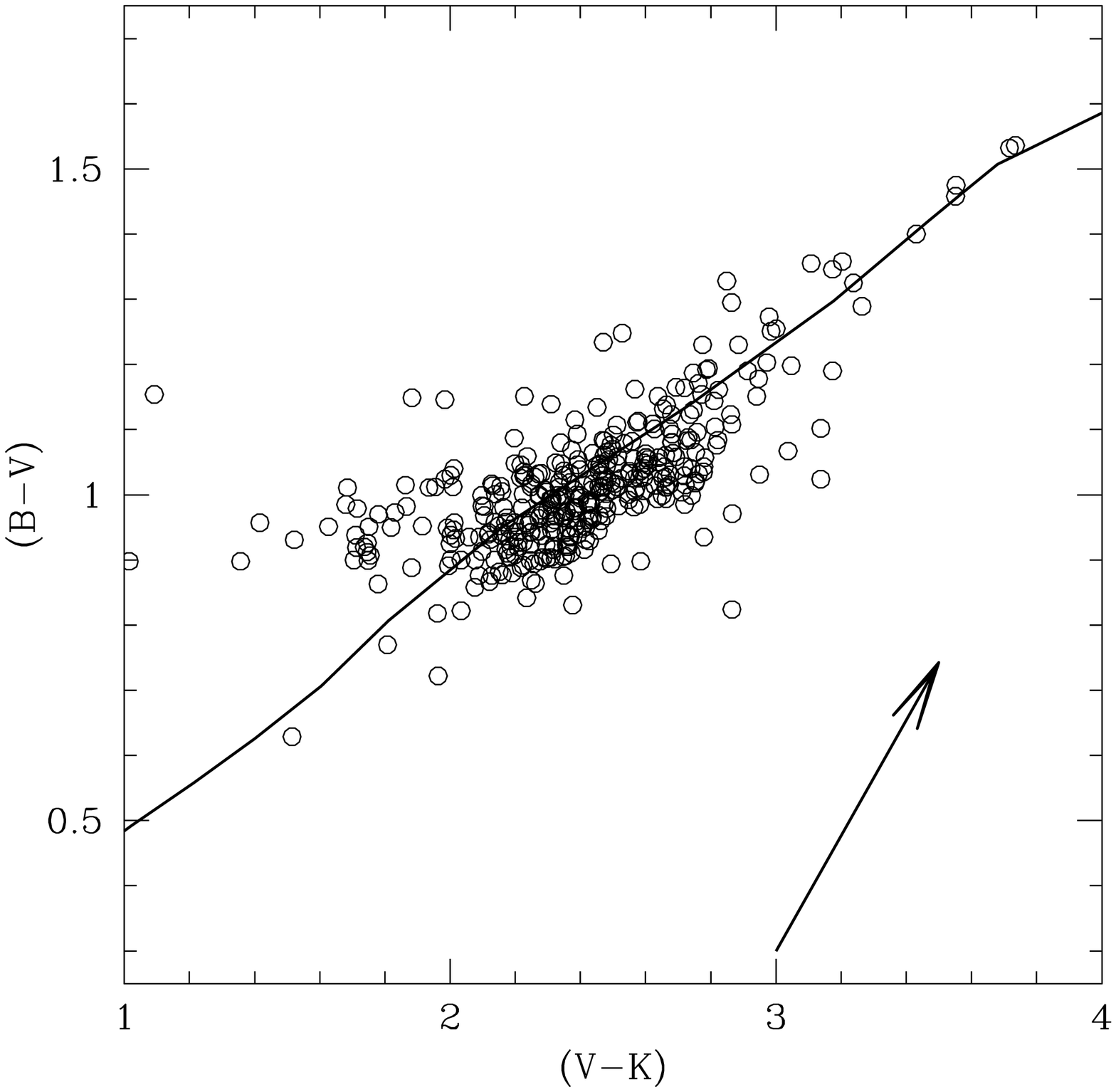,width=9cm,height=9cm}}
\caption{Two color diagram for MS stars in NGC~2141. The solid line is a Zero
Age MS for $[Fe/H]~=~-0.43$. In the lower right corner the reddening vector is shown.
See text for any detail.}
\end{figure}

\begin{figure*}
\centerline{\psfig{file=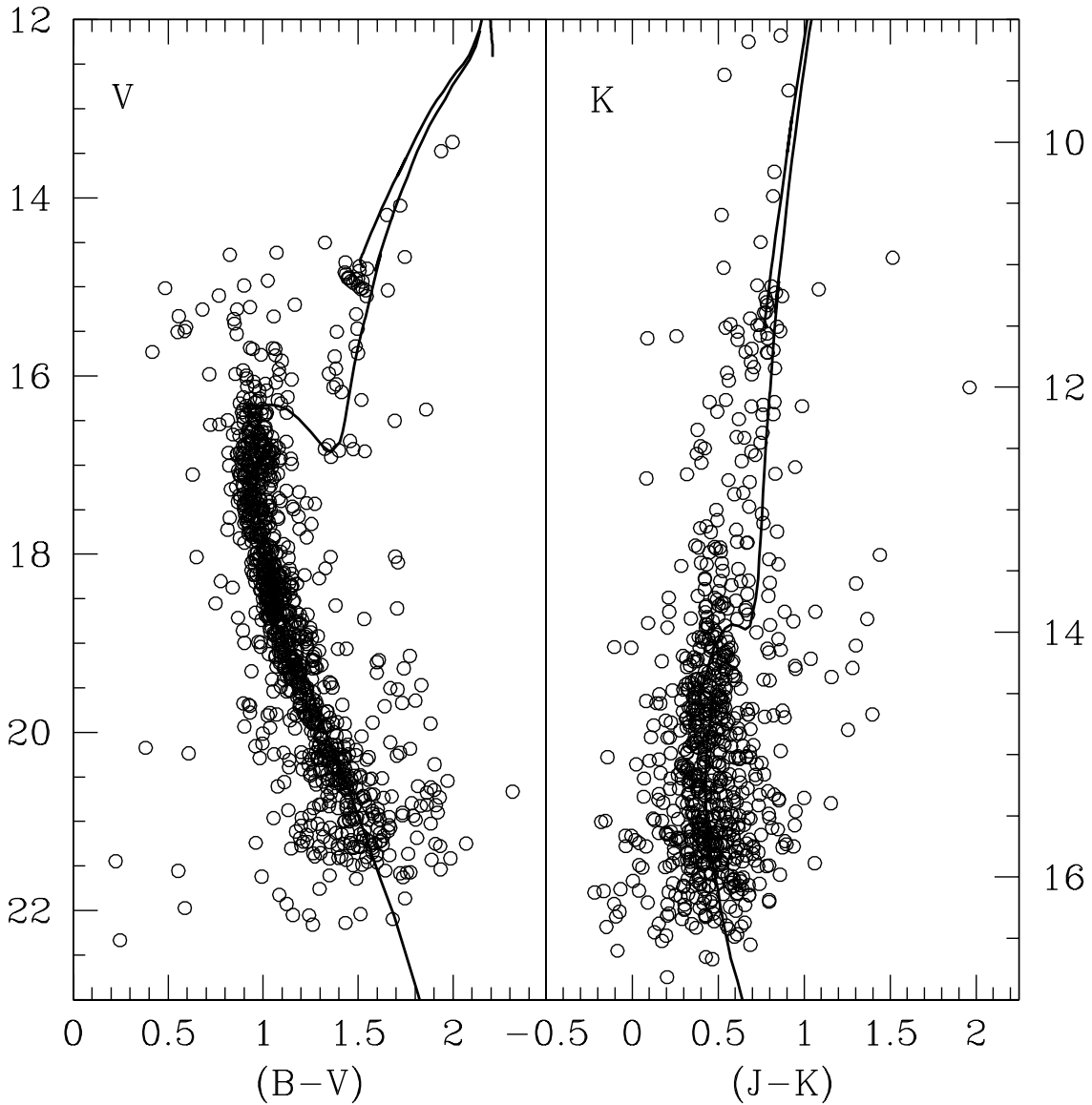,width=16cm,height=16cm}}
\caption{Age determination for NGC~2141. Overimposed on the CMDs
are $Z~=~0.007$ isochrones for an age of $2.5~Gyrs$. The right panel shows
the fit in the plane $K -(J-K)$, whereas the left
panel shows the fit in the plane $V -(B-V))$.
See the text for any detail.}
\end{figure*}

\section{Metallicity}
The CMD diagram in the IR allows us to derive an independent estimate
of the cluster abundance by using a photometric method (Tiede et al. 1997).
This method correlates the slope of
the RGB, defined as $\Delta(J-K)/\Delta K$ with the cluster
metallicity, measured by the index [Fe/H], and for
open clusters, the relation reads:

\begin{equation}
[Fe/H] = -1.639 - 14.243 \times (GB slope).
\end{equation}

\noindent
This method has already been applied by us in the study of NGC~7789 (Vallenari
et al 2000).\\
To find the RGB slope we performed a least squares fit to
the RGB stars. This yields a RGB slope
$\Delta(J-K) / \Delta K$ of -0.085 $\pm$ 0.006.  By using the relation
(3), we obtain [Fe/H] = -0.43 $\pm$ 0.07.  
The reported error is derived as:

\begin{equation}
\Delta([Fe/H]) = 14.243 \times \Delta(GB slope).
\end{equation}

\noindent 
and has to
be considered as an optimistic estimate, since it does not take into
account the uncertainties in the coefficients of eq.~3, and the
sensitivity of the RGB slope to the method adopted for its
computation. However the value we find implies a metal content closer
to the spectroscopic estimate (-0.39 $\pm$ 0.11, Friel \& Janes 1993)
than all the other photometric estimates (Janes 1979, Geisler 1987).

\section{Reddening}
In order to derive the interstellar extinction for NGC~2141, we have combined
optical and IR photometry ($500$ stars in total), and construct two color 
diagrams, namely $(J-K)~vs~(V-K)$ and $(B-V)~vs~(V-K)$, which are
shown in Fig.~5 and Fig.~6, respectively. Only MS stars are considered
to avoid evolutionary effects.
We superimposed a Zero Age MS (ZAMS) for the theoretical
metal content $Z~=~0.007$, obtained translating the observed $[Fe/H]$ by means
of the relation:

\[
[Fe/H] = \log \frac{Z}{0.019}
\]

\noindent
taken from Girardi et al. (2000b).
In Fig.~5 the fit has been obtained by shifting the ZAMS with
E(J-K)~=~0.07 and E(V-K)~=~0.35, which corresponds to a ratio 
$\frac{E(V-K)}{E(J-K)}~=~5.0$, relatively close to the value $5.3$
suggested by Cardelli et al (1989).\\
The fit in Fig.~6, on the other hand, has been achieved by shifting the ZAMS by
E(B-V)~=~0.40 and E(V-K)~=~0.35, whose ratio turns out to
be $\frac{E(V-K)}{E(B-V)}~=~0.86$, in agreement with the value $0.87$
from Cardelli et al (1989).
Although reasonable, these estimates are affected by the limitation that
the reddening vector is almost parallel to the ZAMS,
especially in the $(J-K)~vs~(V-K)$ plane.  This is a minor problem in
the plane $(B-V)~vs~(V-K)$, where we obtain a reddening value
E$(B-V)$= 0.40,  
not much different from E$(B-V)$~=~0.35 derived by Rosvick (1995).
We must stress however that Rosvick (1995)
did not take into account the metallicity
of NGC~2141, which is lower than the Sun.

\begin{figure}
\centerline{\psfig{file=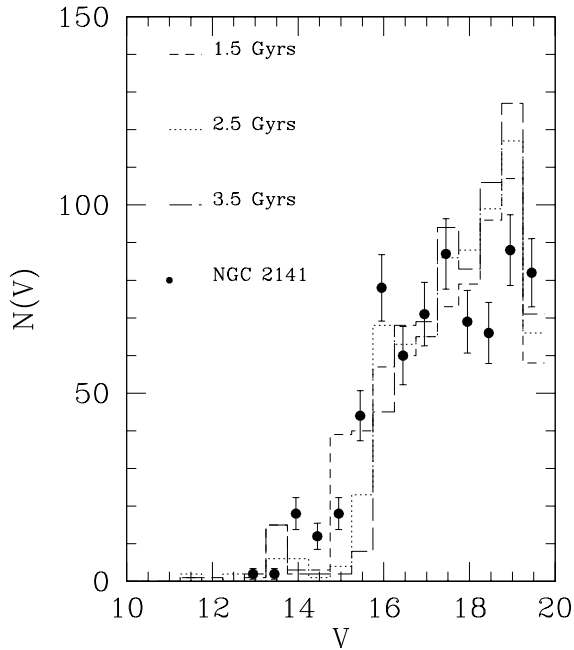,width=9cm,height=9cm}}
\caption{Differential LF of MS stars in NGC~2141 (filled circles). 
Overimposed are three  theoretical LFs 
for $Z~=~0.007$ and ages of $1.5, 2.5$ and $3.5$~Gyrs obtained by assuming
the Kroupa et al (1993) IMF. 
See the text for any detail.}
\end{figure}

\section{Age and distance}
The knowledge of metallicity and reddening, allows us to infer
the distance and age of NGC~2141 by means of fitting with isochrones
(Girardi et al. 2000b).
The fit is shown in the two panels of 
Fig.~7, for the planes $V$ vs $(B-V)$, and $K$ vs $(J-K)$.\\
We have adopted the theoretical metallicity $Z~=~0.007$ derived above,
and performed the fitting with a $2.5~Gyrs$ isochrone, which better
matches the observational data.
The criterion that guided us in performing this comparison was the simultaneous
fit of the TO and the clump magnitudes. Since no membership exists for this
cluster, it is not possible to exactly define the MS TO, which is populated also
by unresolved binary stars, and interlopers. 
By using the reddening estimated derived in the previous Section,
the apparent distance moduli $(m-M)_{K,(J-K)}$ and $(m-M)_{V,(B-V)}$,  
in the plots turn out to be
$13.10$ and $14.15$, respectively. The latter value can also be obtained
assuming that the mean clump magnitude is $M_{V}~=~0.80$ (Girardi et al 1998).
Once corrected, these
values converge to the absolute distance modulus $(m-M)_o~=~12.90 \pm
0.15$. This value is in agreement within the errors with Rosvick (1995) estimate.\\
NGC~2141 turns out to be $3.8\pm0.5$~kpc distant from the Sun, and about 12.0~kpc
far from the Galactic Center.

\section{Luminosity Function}
In this section we compute the differential Luminosity Function
LF in the V 
band . 
The data from the cluster have been previously corrected for
completeness (see Carraro \& Ortolani 1994 for details). 
It resulted 100\% up to $V$~=~19, then
85\% at 19.7, 73\% at 20.2 and 66\% at $V$~=~20.7. Fainter bins
have not been considered because of the large amount of correction.\\

This LF is compared in Fig.~8 with thoretical ones
for the same age range (1.5, 2.5 and 3.5~Gyr) and  metallicity (z=0.007)
of NGC~2141. The theoretical LFs have
been calculated by assuming the IMF proposed by Kroupa et al (1993),
which  is 
somewhat steeper than the classical Salpeter (1955) IMF , 
and best suited for the Solar 
Neighbourhood. It reads:

\[ \begin{array}{l l}
\Phi(M) = &  \left\{
\begin{array}{l l}
C_{k1} \, M^{-0.5}  \, &  \, M < 0.5 \\
C_k \, M^{-1.2}  \, &  \, 0.5 < M < 1 \\
C_k \, M^{-1.7}  \, &  \, M > 1 
\end{array} \right.
\end{array} \]

\noindent
where $C_{k1}~=~0.48$ and $C_{k}~=~0.295$.\\
The comparison clearly shows that the  agreement with theoretical models
is good up to $M_V$~=~5.0 . At fainter luminosities the theoretical
models predict a higher number of stars, increasing with decreasing magnitudes.
This occurs in the main sequence at about 1.0 $M_{\odot}$,
while the TO corresponds to about 1.4  $M_{\odot}$.\\
It seems more reasonable
to interpret these data as a more common mass segregation effect
with low mass stars evaporated from the cluster center.

\section{Conclusions}
In this paper we have presented a detailed study of the poorly
known intermediate age open cluster NGC~2141.
By combining optical and IR photometry we have proved that NGC~2141 is a
moderate age open cluster about $3~Gyr$ old, intermediate in age between the 
NGC~752 and M~67.\\
By studying the color magnitude  and two color diagrams,
we have obtained estimates for the cluster metallicity, reddening
and distance. In detail, we found that the color excesses 
E$(J-K)$ and E$(B-V)$ 
are 0.07, and 0.40, respectively, and that their ratios are in agreement
with the standard values (Cardelli et al 1989).
The derived corrected distance modulus $(m-M)_o~=~12.90$ implies
a distance of 3.8~kpc from the Sun.\\
These findings are supported by the comparisons of the cluster
LF with theoretical ones,
from which we find also evidences of mass segregation below
$M= 1.0 M_{\odot}$.

\begin{acknowledgements}
This study has been financed by the Italian Ministry of
University, Scientific Research and Technology (MURST) and the Italian
Space Agency (ASI).
\end{acknowledgements}

{}

\end{document}